# An Interface for the Virtual Observatory of the University of Guanajuato

René A. Ortega-Minakata[1], Juan P. Torres-Papaqui[1], Heinz Andernach[1]
and Hermenegildo Fernández-Santos[2]
*[1] Departamento de Astronomía, Universidad de Guanajuato*
*[2] Maestría en Medios Interactivos, Universidad Tecnólogica de la Mixteca*
*{rene, papaqui, heinz}@astro.ugto.mx; hfernandez801@gmail.com*

**Abstract**

*We present the first attempts to build a user-friendly interface for the Virtual Observatory of the University of Guanajuato. The data tables will be accessible to the public through PHP scripts and SQL database managers, such as MySQL and PostgreSQL, all administrated through phpMyAdmin and pgMyAdmin. Although it is not made public yet, this interface will be the basis upon which the final front end for our VO will be built.*

*Furthermore, we present a preliminary version of a web front end to the publicly available stellar population synthesis code STARLIGHT (starlight.ufsc.br) which will be made available with our VO. This front end aims to provide an easy and flexible access to the code itself, letting users fit their own observed spectra with their preferred combination of physical and technical parameters, rather than making available only the results of fitting a specific sample of spectra with predefined parameters.*

***Keywords:*** High Performance Computing: applications; Scientific Visualization; Databases: Public Access

## 1. Introduction

Astronomy is nowadays a highly computational science, with astrophysical models being tested numerically with high performance computing, but also with large amounts of new observational and theoretical data produced, analyzed, archived and made available to the public constantly. To cope with this highly complex task, a global network of both astronomers and computer scientists has been formed as the so-called International Virtual Observatory Alliance (IVOA; www.ivoa.net).

In [4] we described the rationale behind the project of the Virtual Observatory of the University of Guanajuato (VO-UG), its implementation strategies and data model. In this paper, we present the first attempts to build a user-friendly interface for our VO, using PHP scripts and SQL database managers, as well as the phpMyAdmin and pgMyAdmin administrators.

We also present a preliminary version of a web front end to the publicly available stellar population synthesis code STARLIGHT (starlight.ufsc.br) which will be made available with our VO. This astrophysical code fits the observed optical spectrum of a galaxy or stellar system with a set of spectra of simple stellar populations (those with a single age and metallicity), assembled from both stellar evolutionary models and libraries of observed stellar spectra. As a result, STARLIGHT infers relevant stellar parameters from the galaxy's spectrum, such as the mass in stars, mean stellar age and mean stellar metallicity, as well as the star formation history of the galaxy. For a detailed description of the technique used by STARLIGHT, see [3].

## 2. A user-friendly interface for the Virtual Observatory of the University of Guanajuato

In its first implementation, our VO service will make public several databases containing diverse astrophysical information for a large sample of galaxies. The user will be able to query the collection of tables in which this information is stored with an SQL search. The tables will be organized in different databases according to their content, and column-by-column descriptions will be made available. A PHP script will allow the user to select a database to query, and direct him to the query page. Figure 1 shows an example of the database selection page and the description page.

All the tables and databases will be administered by either phpMyAdmin or pgMyAdmin. The user will write his/her SQL search in a textbox, which in turn will be run by one of the above administrators within the selected database. Through a PHP script, the results of the query will be dumped to a screen on the service page, and the page will reload. A results file will also be made available to the user through a download button for a limited period of time (15 days), after which the file will be eliminated automatically, using the crontab daemon. Figure 2 shows an example of how phpMyAdmin will help administer databases and tables. Table 1 lists the PHP scripts for the process described in this section.

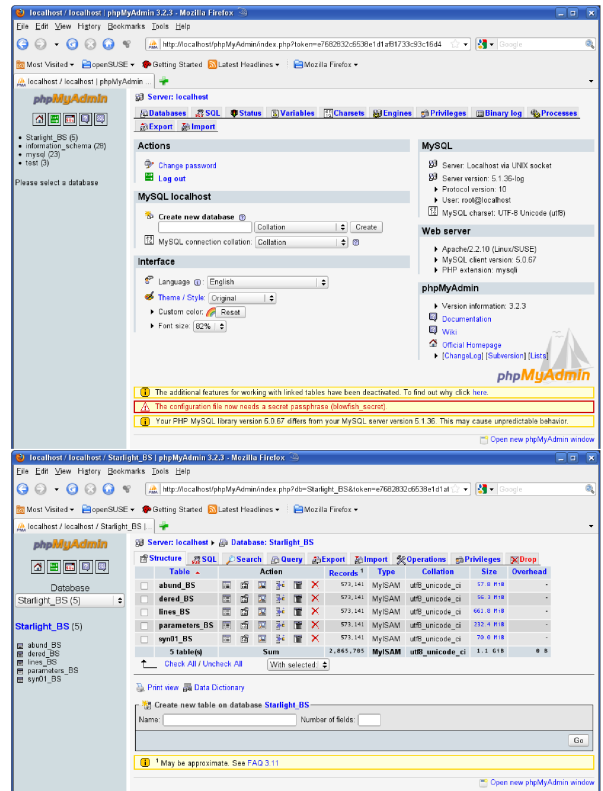

**Figure 2. Example of administration of the different databases *(top)* and the administration of the collection of tables within a single database *(bottom)* using phpMyAdmin.**

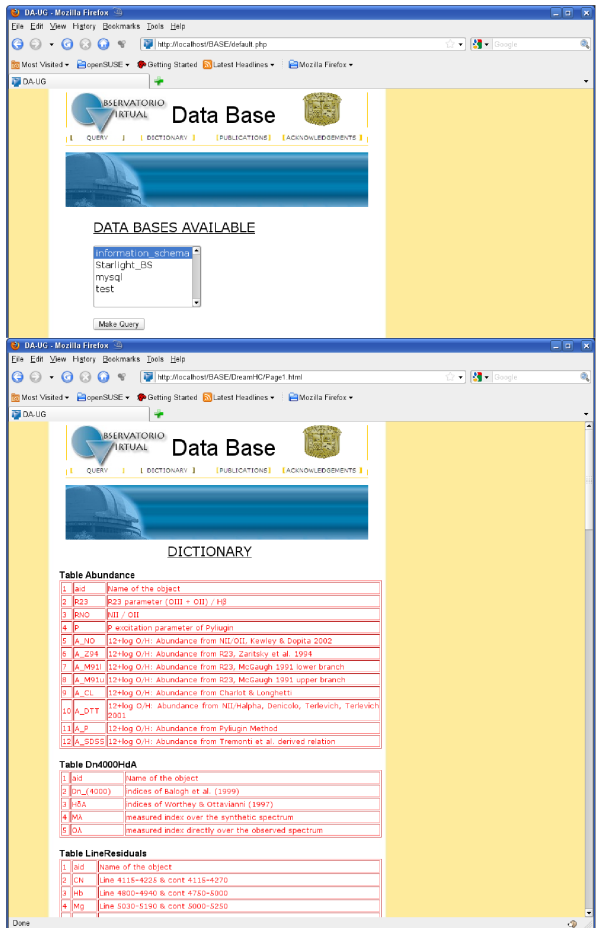

**Figure 1. Example of the database selection page *(top)* and the column-by-column description page *(bottom)* of the Virtual Observatory service to be made available at the University of Guanajuato.**

**Table 1. PHP scripts for the VO database query service.**

| PHP script | Description |
|---|---|
| default.php | Home page for the VO database query service. Lets the user select the database to be queried. |
| search.php | Controls the type of query to be made: a general query (gives the user limited freedom, but no prior knowledge of SQL is required), or a specific query (the user introduces his own SQL query statement in a text box). |
| query.php | Manages the environment to show the query results. |
| specific.php | Executes the query, displays the results, and manages the download button that allows the user to get a results file. |
| download.php | Executes the download of the results file. |
| delete.php | Deletes old user-generated files after 15 days. This script is run automatically using the crontab daemon. |

## 3. A web front end to the STARLIGHT code

In a second phase, our VO will offer a front end to the astrophysical code STARLIGHT. As explained before, STARLIGHT is used to obtain meaningful information from a galaxy or any stellar system by fitting its observed optical spectrum with a set of spectra of simple stellar populations (those with a single age and metallicity), assembled from both stellar evolutionary models and libraries of observed stellar spectra. The information obtained as a result of this fit includes relevant stellar parameters such as the mass in stars, mean stellar age and mean stellar metallicity, as well as the star formation history of the galaxy.

So far, STARLIGHT is available as source code from the STARLIGHT group (starlight.ufsc.br), and the user needs to learn how to use the code and how it works in detail in order to fit their own set of spectra, one by one. The results of applying STARLIGHT to a large sample of galaxies are also available on the same web page (also as a VO service), but these are the results of fits with a predefined set of astrophysical and technical parameters for the whole sample. Our front end aims to provide easy and flexible access to the code itself, letting users fit their own observed spectra with their preferred combination of physical and technical parameters, without the need of learning how the source code works in detail and run the code on each single spectrum.

Through a series of PHP scripts, the user will decide the level of control of the input parameters he desires, depending on his/her knowledge of the involved technicalities (beginner or expert), and upload the files containing the spectra to be fitted. Figure 3 shows the welcome page to the STARLIGHT front end service and the description page for the different modes. On the welcome page, the user informs the service of the number of files (spectra) to be processed, and also chooses the mode (beginner or expert) in which the code will be run.

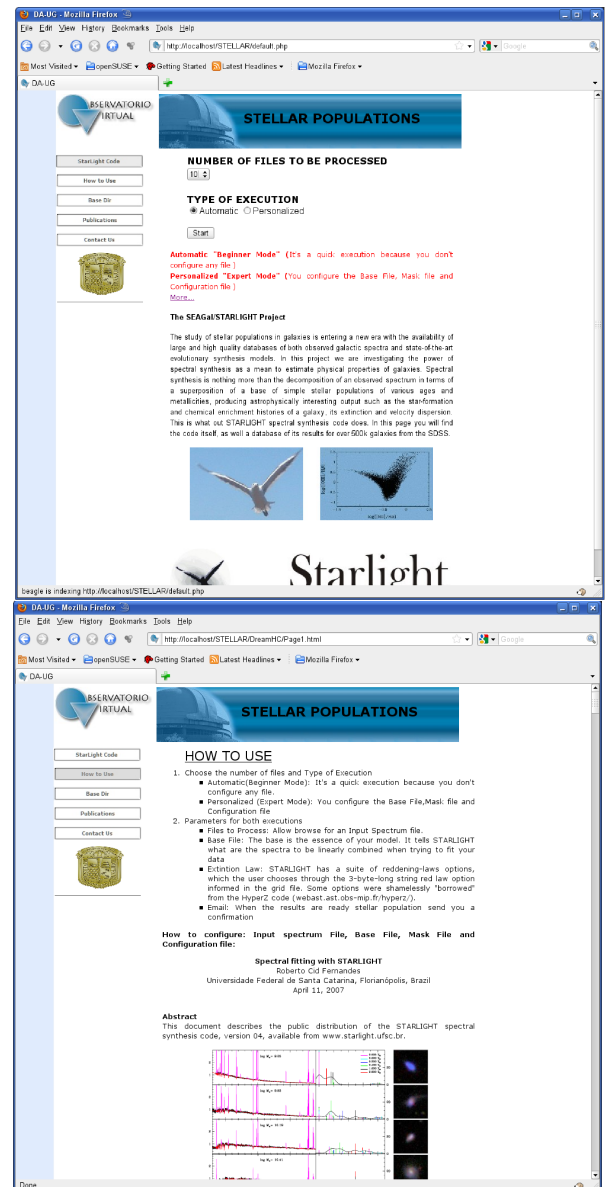

**Figure 3. Welcome *(top)* and description page *(bottom)* of our STARLIGHT front end service.**

Once all spectra are uploaded and configurations specified (depending on the running mode), a PHP script will create a task and add it to a job queue. The task list will be run automatically and periodically using the crontab daemon, and each element on the job queue will be deleted once it has been run. Upon completion of the proper STARLIGHT run, the task will send a confirmation of completion (or detection of errors) e-mail to the user. The user must then return to the web interface and retrieve the result files through a link created by another PHP script. Table 2 lists the PHP scripts for the operation of the STARLIGHT front end described in this section.

**Table 2. PHP scripts for the STARLIGHT front end that will be available at our VO service.**

| PHP script | Description |
|---|---|
| default.php | Home page for the STARLIGHT front-end service. Allows the user to select the running mode and makes the necessary configurations needed for each mode. Also lets the user indicate the number of input files (spectra) to process and manages the upload of these files. It also manages the addition of the job to the queue (task list) and informs the user that the job was submitted. |
| daemon.php | Executes the job queue, dumps the result files to an appropriate directory and sends a confirmation of completion (or detection of errors) e-mail to the user. It also generates the links to download the result files on the web page. |
| download.php | Executes the download of the results files when clickling on the links on the web page. |
| delete.php | Deletes old user-generated files. This script is run automatically using the crontab daemon. |

## 4. Conclusions

The availability of astrophysical data for large samples of objects makes it imperative to find solutions to handle this large amount of information. The standard established by the IVOA marks the direction in which all astronomical data services should move. The Virtual Observatory of the University of Guanajuato will contribute to this task by using a set of standard information technology tools, such as database managers (MySQL, PostreSQL), administrators (phpMyAdmin, pgMyAdmin) and languages (PHP, SQL) to construct a VO service, according to the IVOA standard, to allow the creation of our own astrophysical databases (see, for an example, [1] and [2]).

Furthermore, a user-friendly front end to the STARLIGHT code will help users employ this useful astrophysical code with their own set of technical and physical parameters without the need to learn the specifics of the source code. This flexible and easy access to an otherwise complex code will increase the potential information that can be obtained from using such a code, eventually allowing for an extrapolation of this implementation to similar astrophysical codes.